\documentclass[useAMS]{mn2e}
\usepackage{times}
\usepackage{psfig}
\usepackage{graphicx}

\title[Testing the blazar spectral sequence]{Testing the blazar spectral sequence: X--ray selected blazars}

\author[L. Maraschi et al.]
{L. Maraschi$^1$\thanks{E--mail: laura.maraschi@brera.inaf.it}, L. Foschini$^2$,
G. Ghisellini$^1$, F. Tavecchio$^1$, R. M. Sambruna$^3$ \\
$^1$ INAF -- Osservatorio Astronomico di Brera, V. Brera 28, I--20100 Milano, Italy\\
$^2$ INAF--IASF Bologna, Via Gobetti 101, 40129, Bologna (Italy)\\
$^3$ NASA--Goddard Space Flight Center, Code 661, Greenbelt, MD 20771 (USA)
}

\begin{document}

\pagerange{\pageref{firstpage}--\pageref{lastpage}} \pubyear{2008}
\maketitle

\begin{abstract} 
We present simultaneous optical and X--ray data from \emph{Swift} for
a sample of radio--loud flat spectrum quasars selected from the
\emph{Einstein Medium Sensitivity Survey} (EMSS).  We present also a
complete analysis of \emph{Swift} and \emph{INTEGRAL} data on 4
blazars recently discussed as possibly challenging the trends of the
hypothesised ``blazar spectral sequence''.  The SEDs of all these
objects are modelled in terms of a general theoretical scheme,
applicable to all blazars, leading to an estimate of the jets'
physical parameters. Our results show that, in the case of the EMSS
broad line blazars, X-ray selection does not lead to find sources with
synchrotron peaks in the UV/X-ray range, as was the case for X-ray
selected BL Lacs. Instead, for a wide range of radio powers all the
sources with broad emission lines show similar SEDs, with synchrotron
components peaking below the optical/UV range. The SED models suggest
that the associated IC emission should peak below the GeV range, but
could be detectable in some cases by the Fermi Gamma-Ray Space
Telescope.  Of the remaining 4 "anomalous" blazars, two highly
luminous sources with broad lines, claimed to possibly emit
synchrotron X-rays, are shown to be better described with IC models
for their X-ray emission. For one source with weak emission lines (a
BL Lac object) a synchrotron peak in the soft X-ray range is
confirmed, while for the fourth source, exhibiting narrow emission
lines typical of NLSy1s, no evidence of X-ray emission from a
relativistic jet is found.  We reexamine the standing and
interpretation of the original ``blazar spectral sequence'' and
suggest that the photon ambient, in which the particle acceleration
and emission occur, is likely the main factor determining the shape of
the blazar SED. A connection between SED shape and jet
power/luminosity can however result through the link between the mass
and accretion rate of the central black hole and the radiative
efficiency of the resulting accretion flow, thus involving at least
two parameters.
\end{abstract}


\section{Introduction}

The Spectral Energy Distributions (SED) of blazars from radio to
gamma--rays exhibit remarkable properties, in particular a
``universal'' structure consisting of two broad humps.  A systematic
compilation of multiwavelength data for three complete samples, two
selected in the radio band (2Jy FSRQ and 1 Jy BL Lacs) and one in
X--rays (Einstein Slew Survey BL Lacs) was performed by Fossati et
al. (1998). Note that, at the time, the results from the Compton Gamma-Ray
Observatory (CGRO) were still under study, while very few objects had been 
detected in the TeV band, thus the data at high energies were rather incomplete.

The derived SEDs were averaged in fixed radio luminosity
bins, {\it irrespective of sample membership or classification (BL Lac
vs. FSRQ)}.   The result was
the well known ``Spectral Sequence'' (a series of 5 average SEDs,
``sequence'' in the following) showing the peaks of the two spectral
components shifting systematically to higher frequencies with
decreasing luminosity.  For brevity we will refer to SEDs with
synchrotron peak below or above the optical range as "red" or "blue"
SEDs respectively.

It has to be stressed that the derived average SEDs, by themselves,
only represent the spectral properties of the considered samples and
do not contain further implications; however the fact that, despite
the mixing of different samples and classifications, the average SEDs
show systematic trends, possibly driven by the radio power  {\it
suggests} an underlying regularity of behaviour of jets in very
different objects.
Understanding this behavior could lead to a more profound
comprehension of the jets' radiation mechanisms and physics.

Ghisellini et al. (1998) modelled a large number of individual sources
with sufficient data at high energies with a simple general scheme: a
single emission region, moving with bulk Lorentz factor $\Gamma \simeq
10$, contains relativistic electrons emitting via synchrotron
radiation the lower energy spectral component; the same electrons up
scatter synchrotron photons as well as photons from an external
radiation field producing the high energy spectral component. The
electron spectrum is computed assuming a broken power law injection
spectrum and taking into account radiative energy losses.

With these minimal assumptions the ``sequence'' of spectral properties
translates into a parameter sequence, in which the energy of the
electrons radiating at the SED peaks is higher for lower values of the
(comoving) total magnetic field plus photon energy densities, due to
the reduced energy losses.  This interpretation of the ``sequence''
concept and following developments (Ghisellini, Celotti \& Costamante
2002; Costamante \& Ghisellini 2002) led to successful predictions of
the best candidates for TeV detection (see Wagner 2008 for a recent
review).

On the other hand, the phenomenological ``sequence'' is likely affected by
observational biases (Maraschi and Tavecchio 2001 (ASP CS Vol 227)).  
Significant efforts were made to construct new
samples of blazars with different thresholds and selection criteria
(\emph{e.g.}, Laurent-Muehleisen et al. 1999, Caccianiga \& March\~a
2004, Massaro et al. 2005, Landt et al. 2006, Giommi et al. 2007a,
Turriziani et al. 2007, Healey et al. 2008, Massaro et al. 2008) and
some discrepant objects were found (see Padovani 2007 for a review on
the challenges to the blazar sequence). In particular Caccianiga \&
March\~a (2004) evidenced the existence of sources with relatively low
radio power, but X-ray to radio ratios similar to those of FSRQs.

In a somewhat different line of approach the group lead by
Valtaoja developped a technique to estimate the beaming parameters
(Doppler factors Lorentz factors and viewing angles) of blazar jets
from the radio properties alone (Valtaoja et al. 1999, Lahteenmaki,
Valtaoja \& Wiik 1999, Lahteenmaki \& Valtaoja 1999). Basing on
extensive monitoring data at high radio frequencies (22 and 37 GHz)
these authors compute "apparent" brightness temperatures using light
travel times inferred from variability as sizes for the emission
region.  Comparing the results with the theoretical maximum brightness
temperature allowed by the SSC process they derive the Doppler factor
of the emission region and further use the apparent superluminal
speeds to determine the bulk Lorentz factor and the viewing
angle. This procedure is correct, however one has to keep in mind that
the results obtained refer to the radio emitting regions of the jet,
which do not coincide with the regions emitting the high energy
radiation (e.g. Ghisellini et al. 1998; Sikora 2001).  Moreover in the
analysis of the SEDs of a large number of BL Lac objects Nieppola et
al. (2006) fit the data in the full (radio to gamma-ray) frequency
range available with a single parabolic curve, thus their
determination of the "peak frequency" of the SED cannot be compared
with our approach. In Nieppola et al. (2008) however, the parabolic
fits are not forced to include the X-ray and higher energy data,
yielding peak frequencies that can be compared with ours.  We will
discuss their results in the Discussion Section.

One of the problems in the ``sequence'' assembly was the lack of an
X--ray selected sample of flat--spectrum radio quasars. Wolter \&
Celotti (2001) selected and discussed a sample of FSRQ within the X-ray
selected sample of radio Loud AGN from the
\emph{Einstein Medium Sensitivity Survey} (EMSS). They found that the
average broad--band spectral indices of these X--ray selected FSRQs
were consistent with those of the radio selected FSRQs within the
``sequence''. However, a study of their SEDs was not possible due to
insufficient spectral data.

Here, we present \emph{Swift} observations of the 10 X--ray brightest
objects in the Wolter \& Celotti sample as well as a full analysis of
\emph{Swift} data for 4 blazars recently claimed to be at odds with
the sequence scheme. We model the SEDs of the 14 objects following
Ghisellini, Celotti \& Costamante (2002), deriving estimates of
important physical quantities for their jets. Finally, in the light of
these new results, we discuss the present standing and interpretation
of the blazar ``spectral sequence''.  Preliminary results were given
in Maraschi et al. (2008). The scheme of the paper is as follows:
Sect.~2 gives information on the selected sources and Sect.~3
describes the data and analysis methods.  Sect.~4 details the general
spectral modelling procedure.  The results are presented in Sect.~5. A
general discussion is given in Sect.~6 while  Sect.~7 offers
 the conclusions 

\section{The sources}

\subsection{The X-ray selected Radio Loud AGN sample from the Einstein 
Medium Sensitivity Survey} 
Wolter \& Celotti (2001) selected 39 Radio--Loud, broad line AGN
from the X--ray EMSS sample (Gioia et al. 1990, Stocke et al. 1991). 
Of these, 20 had a measured 
radio spectral index $\alpha_{r} < 0.7$ and their broad band properties
were compared to those of classical radio selected flat
spectrum quasars. Their broad band indices ($\alpha_{ro}$,
$\alpha_{ox}$) were found to be similar to those of FSRQ. 
This is at odds with the case of BL Lac objects, which occupy
very different regions in the $\alpha_{ro}$, $\alpha_{ox}$ plane
depending on whether they derive from X--ray or radio selection. On
the other hand, broad--band indices give only a global information,
which may be ambiguous (see below).  We therefore decided to exploit
the unique capabilities of the \emph{Swift} satellite to observe the
ten X--ray brightest FSRQs of the EMSS sample in order to derive their
X--ray spectrum together with simultaneous optical data allowing a
reliable snapshot of the optical to X-ray SED. 
Observations were performed within the filler program. The data gathered 
and analysis methods are described in Sect.~3.

\subsection{Controversial Sources}

In addition to the EMSS sources, we include in our analysis and
discussion four sources of controversial classifications put forward
by other authors: two ultraluminous, high--redshift, broad line
quasars recently discovered, SDSS~J$081009.94+384757.0$ ($z=3.946$,
Giommi et al. 2007b) and MG3~J$225155+2217$($z=3.668$, Bassani et
al. 2007) have been claimed, on the basis of their broad-band spectral
indices, to show a {\it synchrotron} peak in hard X--rays. If
confirmed, these claims would imply strong violations of the sequence
scheme.

Two other cases concern blazars of intermediate luminosities, namely
RX~J$1456.0+5048$ ($z=0.478$, Giommi 2008) and RGB~J$1629+401$
($z=0.272$, Padovani et al. 2002), which show synchrotron peaks in the
X--ray band and, in addition, exhibit emission lines, weak in the
first source and pronounced, but narrow, in the second one. For these
4 objects, all the existing data obtained with \emph{Swift} and, in
the case of MG3~J$225155+2217$ also the data obtained with
\emph{INTEGRAL} were systematically reanalysed to produce a data set
as homogeneous and complete as possible.

\section{Data analysis}
All the data presented here were analyzed in the same way to guarantee
a homogeneous treatment. The common procedures adopted for the
analysis are described below.

\subsection{Swift}
The data from all three instruments onboard \emph{Swift}, namely BAT,
XRT and UVOT (Gehrels et al. 2004), have been processed and analyzed
with \texttt{HEASoft v. 6.3.2} with the CALDB release of November 11,
2007. The observation log is reported in Tables~\ref{table:obslog1}.

The X--ray Telescope XRT ($0.2-10$~keV, Burrows et al. 2005) data were
analyzed using the \texttt{xrtpipeline} task, selecting single to
quadruple pixel events (grades $0-12$) in the photon-counting
mode. Individual spectra from multiple pointings, together with the
proper response matrices, were integrated by using the
\texttt{addspec} task of \texttt{FTOOLS}. The final count spectra were
then rebinned in order to have at least $20-30$ counts per energy bin,
depending on the available statistics, and spectral fits were
performed with simple power law models and galactic $N_H$. In only
three cases additional absorption or a broken power law fit was
required. The average spectral parameters are reported in
Table~\ref{table:xrt}.

The sensitivity of the hard X--ray detector BAT (optimized for the
$15-150$~keV energy band, Barthelmy et al. 2005) is generally
insufficient to detect extragalactic sources in short exposures,
unless for bright states. Therefore, in order to get meaningful
signal-to-noise ratios we combined all the available pointings for a
given source. All the shadowgrams from pointings of the same source
were binned, cleaned from hot pixels and background (flat-field), and
deconvolved. The intensity images were then integrated by using the
variance as weighting factor. No detection with sufficient
signal-to-noise ratio has been found for any source with the available
exposures and upper limits at $3\sigma$ (corrected for systematic
errors) in two energy bands ($20-40$ and $40-100$~keV) have been
calculated (see Table~\ref{table:hardx}).

Data from the optical/ultraviolet telescope UVOT (Roming et al. 2005)
were analyzed using the \texttt{uvotmaghist} task with source regions
of $5''$ for optical filters and $10''$ for the UV filters, while the
background was extracted from a source-free annular region with inner
radius equal to $6''$ (optical) or $12''$ (UV) and outer radius from
$30''$ to $60''$, depending on the presence of nearby contaminating
sources. In the cases of MS~$0402.0-3613$, SDSS~J$081009.94+384757.0$,
RX~J$1456.0+5048$, and MG3~J$225155+2217$ the presence of a nearby
source prevents the use of a background region with annular shape;
therefore, a circular source free region with $60''$ radius was
selected.  We added a $10\%$ error in flux (corresponding to about
$0.1$~magnitudes) to take into account systematic effects. The summary
of average magnitudes per filters is reported in
Table~\ref{table:uvot}.

\subsection{INTEGRAL}
The hard X--ray ($20-100$~keV) data on MG3~J$225155+2217$ were
obtained from the IBIS/ISGRI detector onboard \emph{INTEGRAL} (Lebrun
et al. 2003). They were analyzed with the \texttt{Off-line Scientific
Analysis (OSA) v. 7.0}, whose algorithms for IBIS are described in
Goldwurm et al. (2003), using the latest calibration files
(\texttt{v. 7.0.2}).

The source was observed serendipitously during the revolutions $316$
(May $15-18$, $2005$) and $337$ (July $17-20$, $2005$) for a total
exposure of $388$ ks.  The results are reported in
Table~\ref{table:hardx}. The joint data from the \emph{INTEGRAL}/ISGRI
together with those of \emph{Swift}/XRT, covering an energy range of
$0.2-100$~keV, can be still fit with a single power-law model with
$\Gamma = 1.41\pm 0.09$ and additional absorption of $N_{\rm
H}^{z}=(2.1_{-1.1}^{+1.3})\times 10^{22}$~cm$^{-2}$, to mime the
low-energy photon deficit.  The fit gives a $\tilde{\chi}^2=0.96$ for
$50$ degrees of freedom.  The results are consistent with the values
for \emph{Swift}/XRT only, reported in Table~\ref{table:xrt}.

\subsection{Two-point spectral indices}
Two point spectral indices $\alpha_{ro}$, $\alpha_{ox}$, $\alpha_{rx}$
between fixed radio/optical/X-ray frequencies have been computed for
all sources, to facilitate comparison with other samples.

The two-point optical/X--ray spectral index has been calculated
according to the formula (Ledden \& O'Dell 1985):

\begin{equation}
	\alpha_{12} = - \frac{\log(S_2 / S_1)}{\log(\nu_2/\nu_1)}
\end{equation}

where $S_1$ and $S_2$ can be the radio ($\nu_r=5$~GHz), optical (at
$\nu_o=8.57\times 10^{14}$~Hz corresponding to $3501$~\AA, $U$ filter)
and X--ray (at $\nu_x=2.42\times 10^{17}$~Hz corresponding to $1$~keV)
flux densities, respectively. The flux densities have been
K--corrected by multiplying for the factor $(1+z)^{\alpha-1}$, where
$\alpha$ is the spectral index of $S_{\nu} \propto \nu^{-\alpha}$. An
average spectral index $\alpha$ has been used for radio and optical
frequencies, with values of $\alpha_r=0.5$ and $\alpha_{\rm o}=1.38$
(Pian \& Treves 1993). The $\alpha_x$ and X-ray flux density have been
measured from the spectral fitting of \emph{Swift}/XRT data reported
in Table~\ref{table:xrt}.

The radio flux densities at $5$~GHz have been extracted from
NED\footnote{http://nedwww.ipac.caltech.edu/} or from radio catalogs
available through
HEASARC\footnote{http://heasarc.gsfc.nasa.gov/}. When the $5$~GHz flux
density was not available (a very few cases), we have extrapolated the
requested value from the $1.4$~GHz flux density.

The \emph{Swift}/UVOT flux densities have been calculated from the
observed magnitudes, dereddened with the $A_V$ from NED and
extrapolated to other frequencies by means of the extinction law by
Cardelli et al. (1989), and converted by using the zeropoints given in
the \texttt{HEASoft} package.

The derived values are reported in Table~\ref{table:uvot}.

\begin{figure*}
\vskip -1.8 true cm
{\hskip -1 true cm
\psfig{file=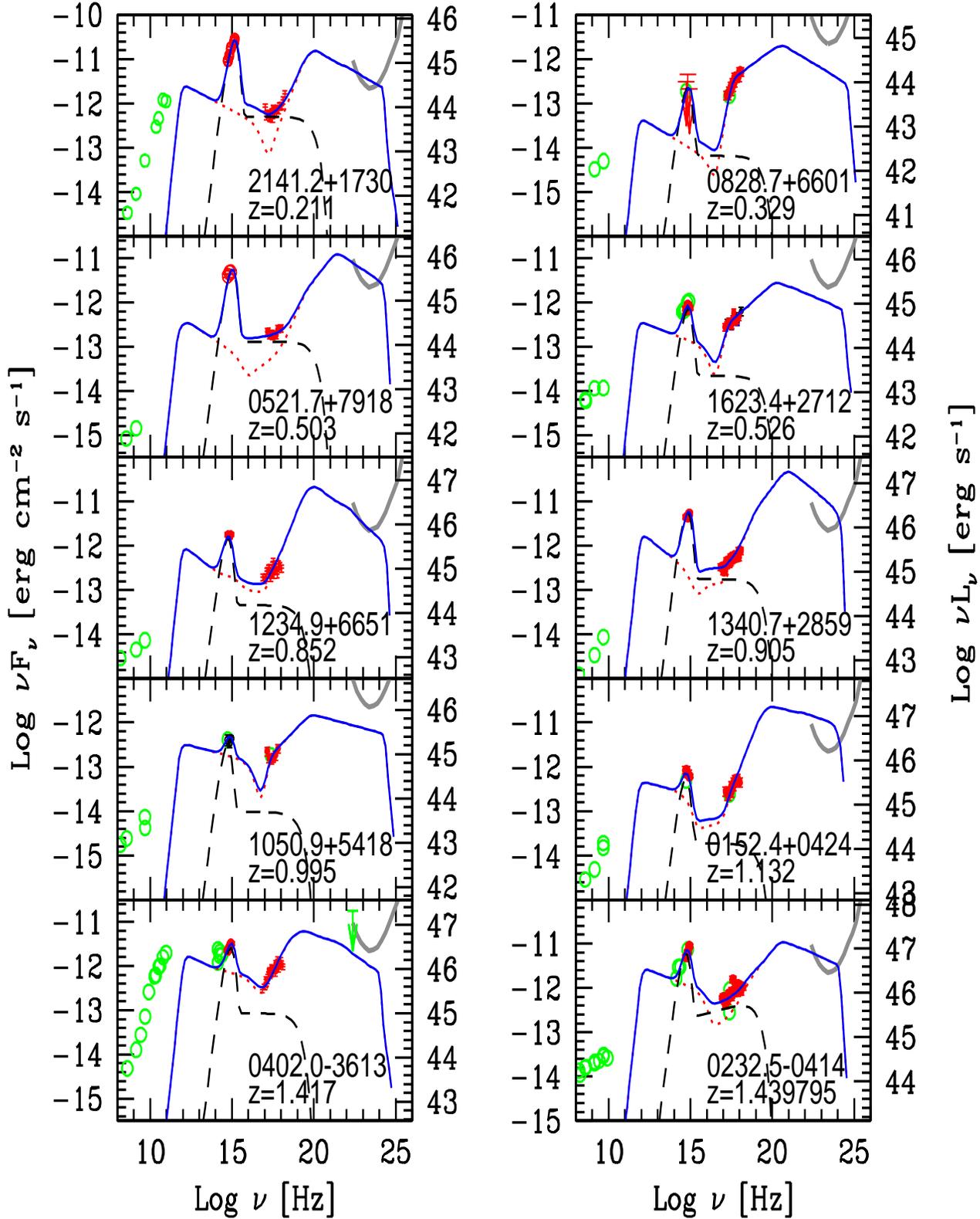,height=25cm,width=18.5cm}
}
\vskip -1.8 true cm
\caption{Spectral Energy Distributions (SED) of the 10 EMSS blazars in
our sample, shown in order of increasing redshift. The models
include the contribution of a putative accretion disk (dashed lines)
described by a blackbody component peaking at $\sim 10^{15}$ Hz plus 
a power law component in X--rays produced by a hot corona. The non
thermal jet emission (dotted lines) is due to synchrotron radiation
(selfabsorbed at mm wavelengths) and IC scattering of photons in
the broad line region (see text). The solid 
lines represent the total emission. Light grey symbols
(green in the electronic version) refer to archival data (taken from
NED), darker symbols (red in the electronic version) represent the
data analyzed in this paper. The gray line indicates the expected 
 flux sensitivity ($5\sigma$, 1 year of exposure) of the FGRT.}
\label{EMSSSED}
\end{figure*}

\section{SED Models} 
\label{modelling}
The observed SEDs of all sources were modelled following the
theoretical scheme developed by Ghisellini, Celotti \& Costamante
(2002) and also used by Celotti and Ghisellini (2008, CG08), where a more
detailed description can be found.

Briefly, the observed radiation is postulated to originate in a single
dissipation zone of the jet, described as a cylinder of cross
sectional radius $R$ and thickness (as seen in the comoving frame)
$\Delta R'= R$.  We assume a conical jet with aperture angle
$\psi_{\rm j}=0.1$, the dissipation is assumed to always occur at
$R_{\rm diss}=10\, R$.

The relativistic particles are assumed to be injected throughout the
emitting volume for a finite time $t^\prime_{\rm inj}=\Delta R'/c$
with total injected power in relativistic particles $L^\prime_{\rm
inj}$ in the comoving frame of the emission region. The observed SED
is obtained from the particle distribution resulting from the
cumulative effects of the injection and radiative energy loss
processes at the end of the injection, at $t=t^\prime_{\rm inj}$, when
the emitted luminosity is maximised.

The finite injection duration, together with the magnetic and
radiation energy densities seen by the particles, determine the energy
$\gamma_{\rm c}$ above which the injected particle spectrum is
affected by the radiative cooling processes.

The particle injection function is assumed to extend from $\gamma=1$
to $\gamma_{\rm max}$, with a broken power--law shape with slopes
$\propto \gamma^{-1}$ and $\propto \gamma^{-s}$ below and above
$\gamma_{\rm inj}$.

If $\gamma_{\rm c} <\gamma_{\rm inj}$ (fast cooling regime) the
resulting particle spectrum $N(\gamma)$ is given by

\begin{eqnarray}
N(\gamma) & \propto & \gamma^{-(s+1)};  
\qquad  \propto \gamma^{-2}; 
\qquad \quad \propto \gamma^{-1}; 
\nonumber\\
\gamma &>& \gamma_{\rm inj};
\qquad \gamma_{\rm c}   <  \gamma < \gamma_{\rm inj};
\qquad \gamma  <  \gamma_{\rm c}
\end{eqnarray}
In the opposite case (slow cooling regime), radiative losses affect
the spectral shape only at higher energies $\gamma > \gamma_{\rm c} >
\gamma_{\rm inj}$, yielding:

\begin{eqnarray}
N(\gamma) &\propto& \gamma^{-(s+1)}; 
\qquad \propto \gamma^{-s};
\qquad \quad \,\, \propto \gamma^{-1}  \nonumber\\
\gamma &>& \gamma_{\rm c};
\qquad \quad \gamma_{\rm inj} <\gamma < \gamma_{\rm c};
\qquad \gamma < \gamma_{\rm inj}.
\end{eqnarray}
The relativistic electrons radiate via the synchrotron and the Inverse
Compton mechanisms. The latter includes the SSC and the external
Compton (EC) processes. The external radiation for the EC process is
assumed to derive from the accretion disk emission scattered and
reprocessed in the surrounding broad line region (Sikora et al. 1994).

With the distributions given above, the synchrotron and IC peaks in
the SED are due to electrons with Lorentz factor, $\gamma _{\rm
peak}$, determined by the injected distribution and the radiative
losses: in the fast cooling regime $\gamma _{\rm peak}=\gamma _{\rm
inj}$ as long as $\gamma _{\rm inj}>1$. When $\gamma _{\rm inj}=1$,
$\gamma _{\rm peak}=\gamma _{\rm c}$ (provided that $s<3$). In the
slow cooling regime, instead, $\gamma _{\rm peak}=\gamma _{\rm c}$ (if
$s<3$).

In most of the sources considered here the data indicate the presence
of a prominent optical/UV excess that one can identify with the disk
emission and then can be used to estimate $L_{\rm disk}$. The disk
emission is assumed to be a black body, peaking at frequency $\nu_{\rm
ext}\sim 10^{15}$ Hz.  Besides this black body UV component, we also
consider the emission of the X--ray corona.  This X--ray component is
assumed to have $\alpha_X=1$, to start a factor 30 below the UV bump,
and to have an exponential cutoff at 150 keV. When no optical/UV
excess is apparent we still include a possible accretion disk (that
can be considered as an upper limit). The external radiation field is
derived assuming for the Broad Line Region a luminosity $L_{\rm
BLR}=0.1 L_{\rm disk}$ and a size $R_{\rm BLR}$ derived from the
relation %
\begin{equation}R_{\rm BLR} \, \sim \,  10^{17} L_{\rm disk, 45}^{1/2}\,\,\, {\rm  cm},\label{kaspi}\end{equation}%
where $L_{\rm disk}= 10^{45} L_{\rm disk, 45}$ erg s$^{-1}$, in
approximate agreement with Kaspi et al. (2007). Since $U_{\rm BLR} =
L_{\rm BLR}/(4\pi R_{\rm BLR}^2 c)$,
the radiation energy density of the BLR
photons, as measured by an observer inside the BLR at rest with
respect to the black hole, is constant, and equal to $U_{\rm BLR} \sim
0.026$ erg cm$^{-3}$. In the comoving frame, however, the radiation
energy density will be enhanced by a $\Gamma^2$ factor.  When $R_{\rm
BLR}$ is smaller then the dissipation zone $R_{\rm diss}$, we
completely neglect any external radiation.

The essential parameters of the model and their role can summarized as
follows:

\begin{itemize}

\item $R$, $L^\prime_{\rm inj}$, $s$ and $\gamma_{\rm inj}$ determine
the injected relativistic particle distribution;

\item the magnetic field $B$ and the radiation energy density $U_{\rm
rad}$ determine $\gamma_c$, the energy of the cooling break, and thus
the particle spectrum after cooling. $U_{\rm rad}$ includes the
contribution of both the synchrotron photons and the external photons.
In the comoving frame $U^\prime_{\rm ext} \sim \Gamma^2 U_{\rm ext}$,
therefore also the bulk Lorentz factor $\Gamma$ is involved.

\item the viewing angle $\theta$, together with $\Gamma$, determines
the Doppler (beaming) factor $\delta$, and then the boosting of the entire SED;

\item the ratio $\Gamma^2 U_{\rm ext}/B^2$ determines the EC to
synchrotron luminosity ratio.

\end{itemize}

The SED peak frequencies ($\nu_{\rm peak,S}$, $\nu_{\rm peak,IC}$)
depend mainly on the shape of the particle spectrum. If $\gamma_c$ is
high (low cooling) $\gamma_{\rm peak} > 10^4$ and $\nu_{\rm peak,S}$
is above the optical band and the SSC dominates the high energy
emission. If $\gamma_c$ is low (strong cooling) $\gamma_{\rm peak} <$
10--100 and then $\nu_{\rm peak,S}$ is very much below the optical band,
often below the self--absorption frequency, and the EC mechanism
produces the high energy component.

Finally, the power carried by the jet in different forms (radiation,
magnetic field, relativistic electrons, cold protons) can be computed
from the model parameters as:

\begin{equation}
L_{\rm i} \simeq \pi R^2 \Gamma^2 \beta c U_{\rm i}
\end{equation}

\noindent where $ U_{\rm rad} = U'_{\rm rad}$ , $ U_{\rm B} = B^2/{8\pi}$, $ U_{\rm e} = n_e <\gamma> m_e c^2$, $ U_{\rm p} = n_p m_p c^2$

We refer to CG08 for a recent estimate and
discussion of the jet powers for a large number of blazars.

\section{Results}
\subsection{The EMSS sample}
In all cases the X-ray spectra are well fit by simple power laws with
photon indices ranging from 1.8 to 1.3 and the $\alpha_{ox}$ index,
which indicates the relative intensity of the X-ray to optical
emission, varies from 0.9 to 1.6.
None of the objects is detected in the hard X--rays with BAT.

In radio--quiet AGN, where the optical as well as X--ray emission
derive from an accretion disk, this ratio is on average $\alpha_{ox}
\simeq 1.6$ (see, e.g., Vignali et al. 2003). Thus, the lower values
of $\alpha_{ox}$ for the EMSS sample, i.e. the high ratios of X--ray
to optical fluxes (see Table~\ref{table:uvot}) point to a significant
contribution from the jet in the X--ray band. This conclusion is
reinforced by the derived spectral indices in the X--ray band
($\alpha_{x} \simeq 0.3 - 0.7$) which are unusually hard for normal,
even radio loud but not flat spectrum, quasars ($\alpha_{x} \simeq
0.8-1 $, see Grandi et al. 2006, Guainazzi et al. 2006).  Therefore,
the observed X-ray emission can be attributed to the combined
contributions from the accretion disk corona plus a harder X-ray
component deriving from Inverse Compton scattering of external photons
(EC) off relativistic electrons in the jet. We recall that all the
objects in the sample have Broad Line Regions providing seeds for the
EC process.

The optical and X--ray fluxes derived from \emph{Swift} for the $10$
X--ray selected radio-loud AGN are shown in Fig.~\ref{EMSSSED} (in
order of redshift), together with historical radio and optical data,
and theoretical models for the full SEDs. The high optical/UV fluxes
and hard X-ray spectra indicate that the UV to X-ray SED have a
concave shape as commonly observed for FSRQs.  Therefore, in all
cases the synchrotron peak frequency should fall below the UV range
(red SED). On the other hand, the lack of data in the IR-submm region
does not allow to set stronger observational constraints on its
location.

It is noteworthy that the two objects with the highest $\alpha_{ox}$,
MS~$2141.2+1730$ and MS~$0521.7+7918$, have relatively low radio
luminosity ($\approx 10^{42}$~erg s$^{-1}$). This suggests that the
jets in these sources could be either intrinsically weak or seen at
intermediate angles, therefore not strongly beamed (see also Landt et
al. 2008). Indeed, the unified model predicts that blazars at
intermediate angle should exist (in large numbers). Sources with
subluminous jets were not present in the samples used to construct the
sequence, likely because the high threshold in radio flux selected
only the most beamed sources.

The X--ray selected AGN considered here, whose radio emission was in
some sense measured ``a posteriori'', have radio fluxes on average
around $200$~mJy and radio luminosities in the range $10^{42-44}$~erg
s$^{-1}$, about one order of magnitude lower than the range covered by
the FSRQs in the sequence. The optical fluxes, indicative of the
accretion disk emission, correspond instead to luminosities of
$10^{45-47}$~erg s$^{-1}$, well in to the typical quasar range. An
even lower radio power interval was explored by Caccianiga and Marcha
(2004), though without spectral information.

MS~$0828.7+6601$ is extreme in the opposite way, in that the optical
luminosity is lower than the X--ray one and the X--ray spectrum is
extremely hard ($\alpha_{x}=0.3$). The Galactic absorption column is
$N_{\rm H}=4.31\times 10^{20}$~cm$^{-2}$ and the X--ray spectrum does
not show any hint of additional absorption, suggesting that the source
is intrinsically faint at optical wavelengths.

The SED models shown in Fig. 1 (continuous/blue lines) were computed
as described in Section 3. As noted above the main observational
constraints are that the synchrotron peak should fall below the UV
range and that the jet contribution in X-rays should have a hard
slope.  Observational data below the optical and above the X-ray range
are lacking so that the models cannot be more strongly constrained.
According to our SED models the peak of the high energy emission,
always dominated by the EC process, falls around 1-10 MeV, unusually
low compared to known $\gamma$--ray blazars.  $\gamma$--ray
observations would be extremely important to confirm these models: the
sensitivity curve of the Fermi GRT,for a 1 yr exposure, shown for
comparison on each SED, indicates that several of these sources may be
detected in the near future.

The model parameters adopted are reported in Table \ref{table:para}.

\begin{figure}
\centering
\includegraphics[scale=0.4]{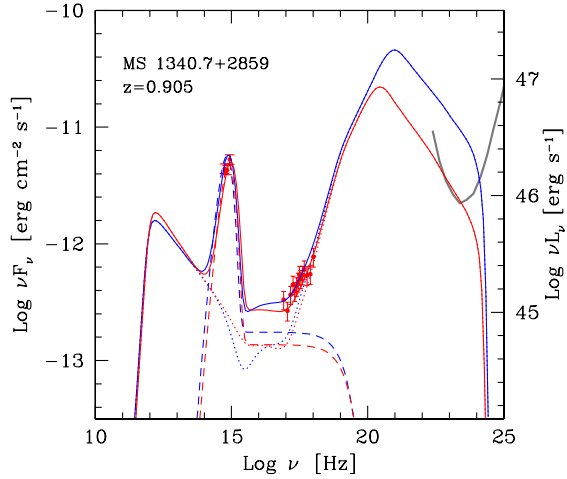}
\caption{Spectral Energy Distributions of MS~$1340.7+2859$, together
with two possible fits. The one with the high energy emission above
the GLAST detection curve (blue line), is the same as in
Fig. \ref{EMSSSED}. The parameters for the two fits, listed in
Tab. \ref{table:para}, differ only slightly, but the predicted flux in
the GLAST band is quite different.}
\label{1340}
\end{figure}
%

Since the lack of  high energy (hard X--ray and $\gamma$--ray) data leaves 
some freedom in the choice of the fitting parameters we present also 
examples of alternative models for the SEDs. 

In Fig.~\ref{1340} two models for the SED of MS~$1340.7+2859$, both fitting
the optical and X--ray data equally well, are compared. 
The first (same as in Fig.~\ref{EMSSSED}) predicts a $\gamma$--ray flux well
above the FermiGRT sensitivity; the second predicts a lower $\gamma$--ray 
flux, just below the FermiGRT  sensitivity limit. 
The parameters for the two models are quite similar 
(see Tab. \ref{table:para}). 

\begin{figure}
\centering
{\hspace*{-1.45 truecm}
\psfig{file=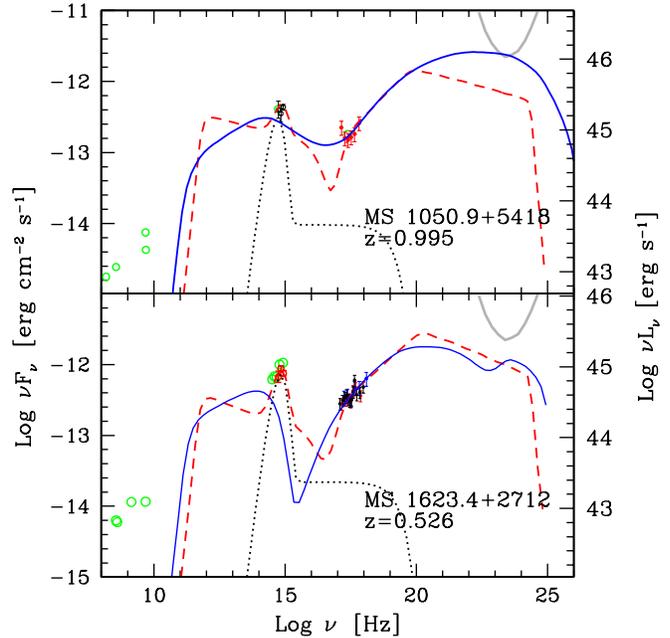,height=10.3cm,width=10.3cm}}
\vspace*{-0.5 truecm}
\caption{SED of MS~$1050.9+5413$ and MS~$1623.4+2712$ together with
two possible fits. For both sources the dashed line is the same fit as
in Fig. \ref{EMSSSED}. The solid line, instead, corresponds to
assuming that the emission region is beyond the radius of the broad
line region. Consequently, the EC process is completely
neglected. This corresponds to less radiative cooling and a larger
$\gamma_{\rm peak}$, and in turn this corresponds to a synchrotron
peak at much larger frequencies. In this case, the high energy flux is
completely due to the SSC process.}
\label{ec_ssc}
\end{figure}

In Fig. \ref{ec_ssc} alternative SED models (solid lines) for
MS~$1050.9+5413$ and MS~$1623.4+2712$ are shown. Both sources are
characterized by a relatively small accretion disk luminosity, hence a
small $R_{\rm BLR}$, a ``normal" X--ray to optical luminosity ratio
and a ``normal" X--ray slope (as opposed to e.g. MS 0828.7+6601, which
has a relatively stronger, and harder, X--ray component). We
hypothesize here that the dissipation region of the jet is beyond the
BLR (i.e. $R_{\rm diss}>R_{\rm BLR}$), contrary to what assumed for
the models shown in Fig.~\ref{EMSSSED}. Then, in the absence of
external photons, the cooling threshold $\gamma_c$ is high and the
synchrotron emission peaks close to (but not above) the optical band,
while the whole high energy emission from X-rays to $\gamma$-rays is
produced by the SSC (instead of EC) process. Note that the injection
spectrum is also different from the previous models, in that
$\gamma_{\rm inj}$ changes from 1 to 400 and 300 for the two sources
respectively (see Tab.~\ref{table:para}). As a consequence the jet
powers computed for the latter models are significantly reduced
(Tab.~\ref{table:power}).

These examples illustrate the ambiguities in our models when
$\alpha_x$ and $\alpha_{ox}$ are ``normal", and the accretion disk
luminosity is relatively modest, suggesting a small broad line
region. It is clear from Fig. \ref{ec_ssc} that additional data,
especially between the sub--mm and the optical region of the SED are
needed to discriminate between these alternatives. In the
following we discuss the results assuming in all the cases the
parameters obtained with the standard external Compton model.

\subsection{Controversial blazars}

\subsubsection{Two high red-shift ``red'' blazars}

\begin{figure}
\centering
{\hspace*{-0.8 truecm}
\psfig{file=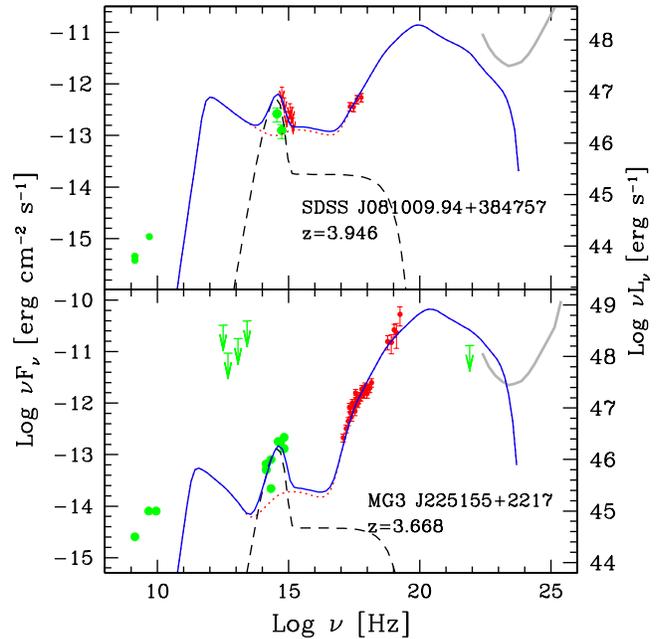,height=10.3cm,width=10.3cm}}
\vspace*{-0.5 truecm}
\caption{Spectral Energy Distributions (SED) of the 2 high--redshift
blazars: MG3~J$225155+2217$ and SDSS~J$081009.94+384757.0$. Same
symbols as in Fig. \ref{EMSSSED}.}
\label{sedhighz}
\end{figure}

Two high redshift powerful blazars recently discovered
(SDSS~J$081009.94+384757.0$, Giommi et al. 2007b; MG3~J$225155+2217$,
Bassani et al. 2007, Falco et al. 1998), both showing conspicuous
broad lines, have been claimed to possibly exhibit a {\it synchrotron
peak} in the X--ray band at variance with the sequence scheme. Our
reanalysis of all the existing \emph{Swift} and \emph{INTEGRAL} data
is summarized in the Tables and the results are shown in
Fig.~\ref{sedhighz}.  For SDSS~J$081009.94+384757.0$ the \emph{Swift}
observations yield only upper limits in the optical UV
range. Therefore the optical fluxes measured by the SDSS, though not
simultaneous to the X-ray data, are also shown in
Fig.~\ref{sedhighz}. In both cases the hard X-ray spectra suggest a
concavity of the optical to X--ray SEDs, pointing to an
inverse-Compton origin of the X--ray/hard X--ray emission. Though both
objects have $\alpha_{RX} \simeq 0.6$, less than the conventional
threshold value of 0.78 taken to distinguish HBL from LBL, the "red"
nature of their SEDs seems clear even with these limited data.

The case of MG3~J$225155+2217$ discovered with \emph{INTEGRAL} is
extraordinary: the X--ray to hard X--ray component dominates the
optical emission by more than one order of magnitude. Attributing the
latter component to synchrotron emission would require implausibly
high values of both magnetic field and particle energies. The SED of
this source is very similar to that of Swift~J$0746.3+2548$
($z=2.979$) discovered by the BAT instrument on board \emph{Swift}
(Sambruna et al. 2006a).

Clearly, a hard X--ray selection favors the discovery of objects with
an extremely dominant hard X--ray component. It is nevertheless
interesting that blazars with such extreme dominance of the high
energy emission component actually exist.

Our model SEDs are shown in (Fig.~\ref{sedhighz}) For
SDSS~J$081009.94+384757.0$, although the wavelength coverage is poor,
the hard X-ray spectrum measured by Swift strongly disfavors both a
synchrotron or SSC interpretation of the X-ray emission. We therefore
propose a model with synchrotron peak at low energies and a strong EC
component. On the other hand the peak energy of the EC component is
uncertain as in the case discussed above for MS1340.7.

For MG3~J$225155+2217$ the INTEGRAL data define well the lower energy
side of the EC component: this branch of the SED reveals the low
energy end of the electron distribution. The SED model is
sufficiently determined by the data and the peak energy of the EC
component is unlikely to be higher.

\begin{figure}
\centering
{\hspace*{-1.45 truecm}
\psfig{file=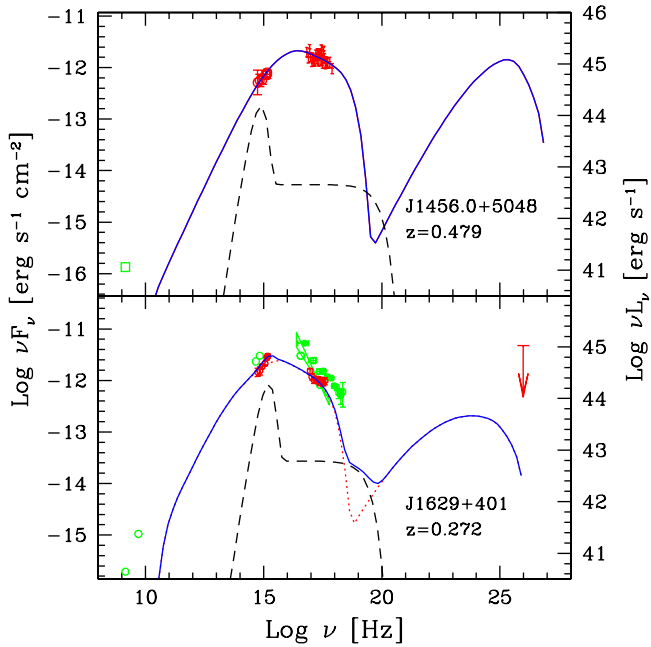,height=10.3cm,width=10.3cm}}
\vspace*{-0.5 truecm}
\caption{Spectral Energy Distributions (SED) of the 2 ``blue'' blazars
with emission lines, RX~J$1456.0+5048$ and RGB~J$1629+401$. The
plotted disk and corona spectra should be considered as an upper limit
to the thermal emission in these blazars. Since the dissipation region
is beyond the location of the BLR, their photons so not contribute to
the inverse Compton spectra.}
\label{sedunusual}
\end{figure}

\begin{figure}
\centering
\includegraphics[scale=0.25]{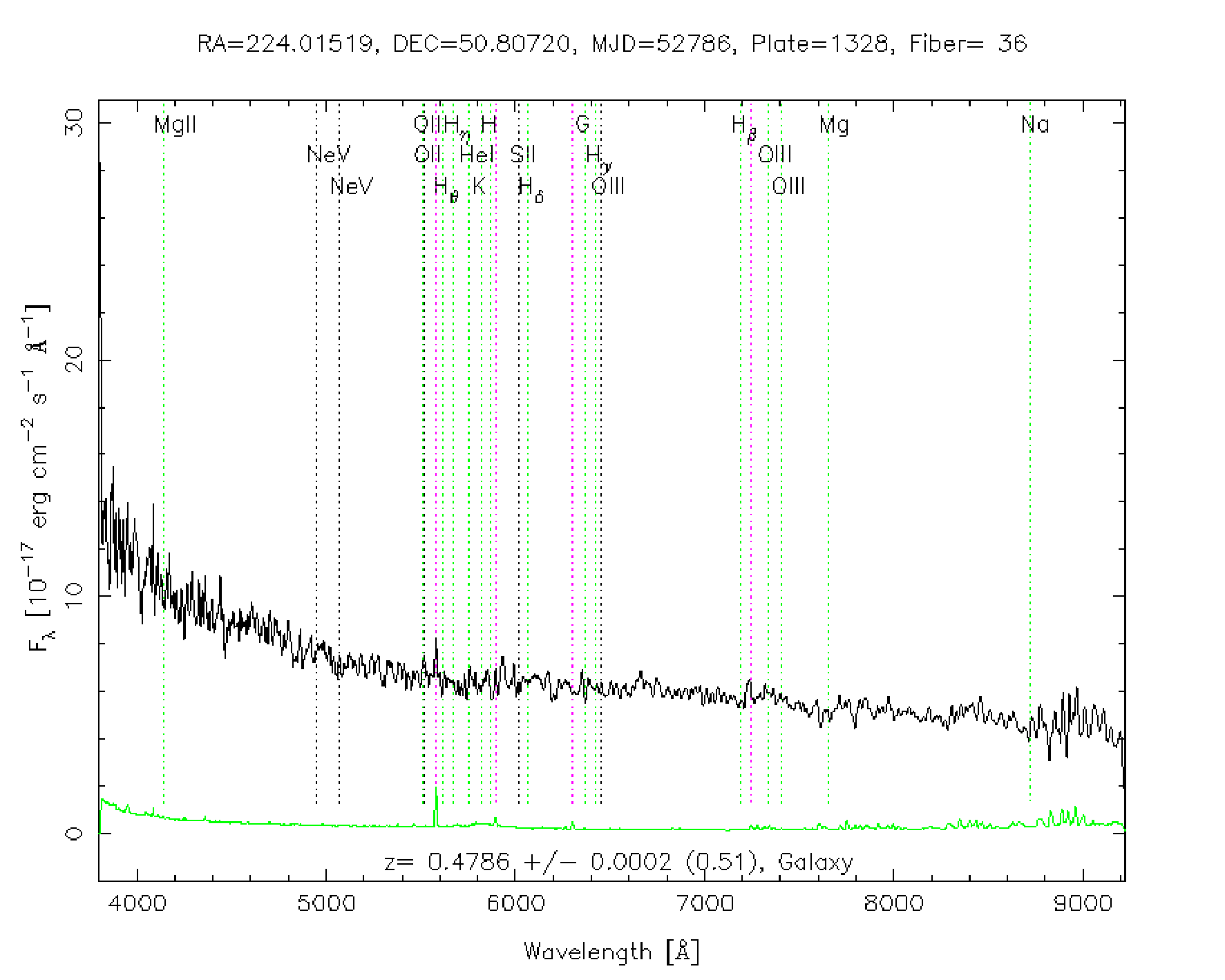}
\caption{Optical spectrum of RX~J$1456.0+5048$ from the Sloan Digital Sky Survey (SDSS).}
\label{sdss1456}
\end{figure}

\begin{figure}
\centering
\includegraphics[scale=0.25]{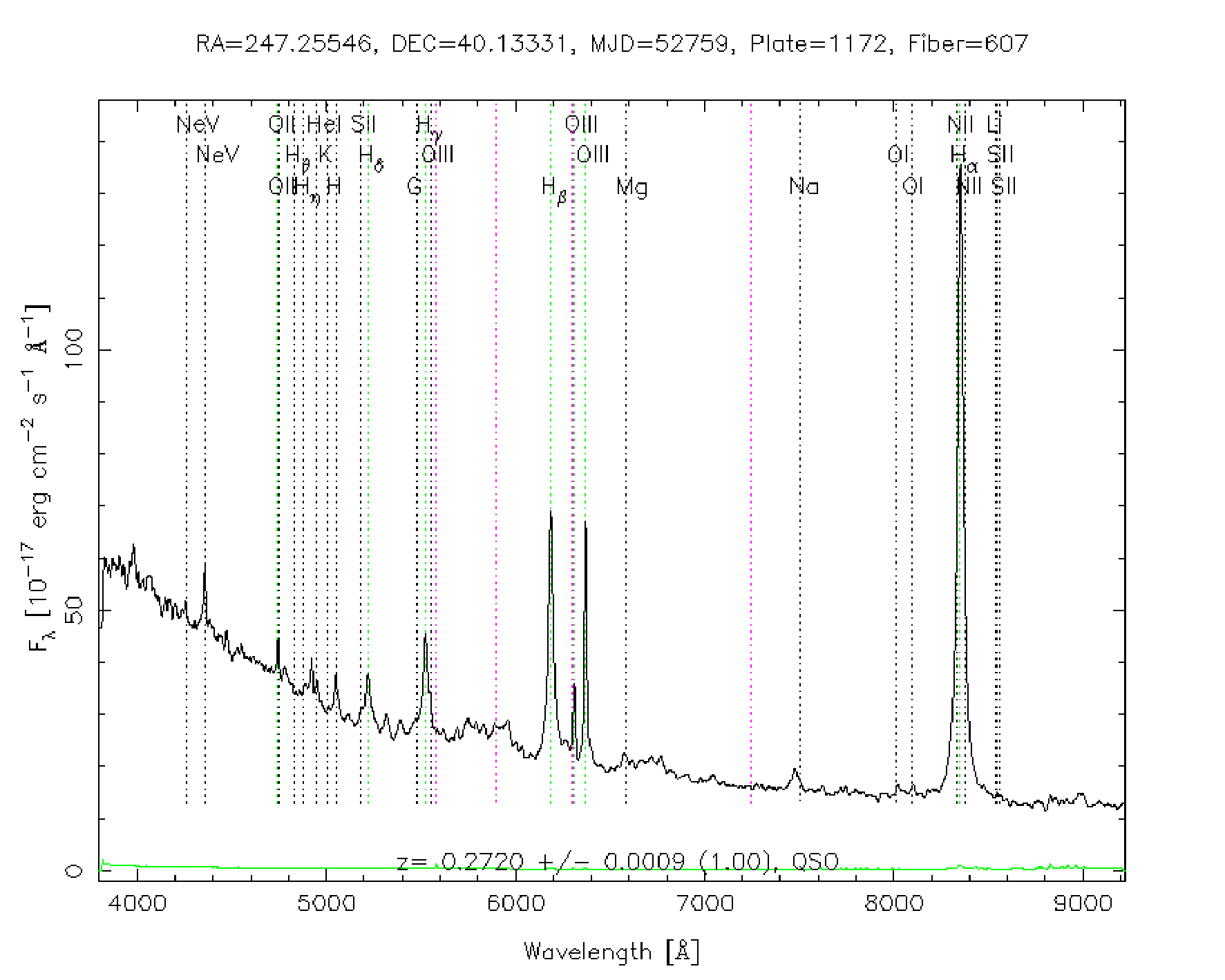}
\caption{Optical spectrum of RGB~J$1629+401$ from the Sloan Digital Sky Survey (SDSS).}
\label{sdss1629}
\end{figure}

\subsubsection{Two luminous ``blue'' blazars}

The SEDs for RX~J$1456.0+5048$ and RGB~J$1629+401$ including all the data 
analysed here, are shown in
Fig.~\ref{sedunusual}. Here the data confirm the ``blue'' nature of
the SEDs as discussed already in Padovani et al. (2002) and Giommi
(2008) respectively.  In both cases the synchrotron peak falls in the
UV to X-ray range as typical for "High energy peaked" BL Lacs (HBL).

Indeed RX~J$1456.0+5048$ can be classified as a BL Lac, since its
optical spectrum from the SDSS (see Fig.~\ref{sdss1456}) shows a blue
continuum and very weak emission lines, with equivalent width
$<5$~\AA\footnote{According to the line measurements available in the
SDSS server at http://cas.sdss.org/astrodr6/en.}  We note also that
the radio source previously associated with this object is probably a
misidentification: a bright, compact flat-spectrum radio source in the
field, about 44 arcsec to the east of the SDSS BL Lac object,
dominates the radio flux in this region. In the best image from the
FIRST, at 1.4 GHz, the field source is 168 mJy, whereas the BL Lac
object is only 9.5 mJy (T. Cheung, private communication). The latter
value for the radio flux is shown in Fig.~\ref{sedunusual}.
RX~J$1456.0+5048$ is highly luminous, but still comparable to the well
known HBL PKS~$2155-304$ in the bright state recently observed
(Foschini et al. 2007, 2008).

The SED model for this object shown in Fig.~\ref{sedunusual} formally
assumes $R_{\rm diss}>R_{\rm BLR}$, however there is no indication
that a broad line region exists at all in this object. Also the
possible contribution of an accretion disk is only an upper limit. The
cooling threshold $\gamma_c$ is very high, as well as the break energy
of the injection spectrum $\gamma_{\rm inj}$ which coincides with
$\gamma_{\rm peak}$.

The case of RGB~J$1629+401$ is intriguing. Its optical spectrum, also
from the SDSS, is shown in Fig.~\ref{sdss1629}. The emission lines are
pronounced, but ``narrow'' ($FWHM < 1500$~km/s, Komossa et al. 2006),
typical of Narrow-line Seyfert 1 galaxies.  Estimates of the mass of
the central BH are in the range $2 \times 10^7$ M$_{\odot}$ (Komossa
et al. 2006).  This is unusual for a radio loud source and in
particular for blazars. However mass estimates for NLSy1s may have to
be revised (Marconi et al. 2008, Decarli et al. 2008). In fact the
whole case of NLSy1 and particularly that of radio loud NLSy1 is the
subject of active debate (e.g. Malizia et al. 2008, W. Yuan et
al. 2008). Specifically RGB~J$1629+401$ is only moderately radio loud
and its X-ray emission spectrum showing a broken power law spectrum
with spectral indices $\Gamma_1 = 3.4^{+0.7}_{-0.3}$ and $\Gamma_2
=2.21\pm 0.06 $ below and above 0.5 keV does not point unambiguously
to a jet contribution.  The X-ray emission could be driven by its
NLSy1 nature (Komossa et al. 2006)

Nevertheless we modelled the SED of this source assuming a jet contribution
and negligible radiative losses in the dissipation region ($R_{\rm
diss}>R_{\rm BLR}$). The radiation energy density is thus due to magnetic
field and synchrotron photons only. The resulting SSC component has low
luminosity.

The parameters derived modeling the observed SEDs for all the objects
are given in Table~\ref{table:para}. The powers carried by the jets in
various forms, derived from the models, are reported in
Table~\ref{table:power}.

\section{Discussion}
\subsection{Model parameters and correlations}

It is interesting to compare the (model independent) two point
spectral index $\alpha_{ox}$ with quantities derived from the
modelling for the EMSS sample.  $\alpha_{ox}$ describes the ratio of the
 optical to X-ray emission. For the ``red'' SEDs of this sample the 
optical is often
dominated by the accretion disk emission while the X-ray contribution in
excess of the accretion disk corona can derive from
the jet, therefore $\alpha_{ox}$ is a measure of their relative
strength.  

\begin{figure}
\centering
{\vspace*{-1.5 truecm}
\hspace*{-1. truecm}
\psfig{file=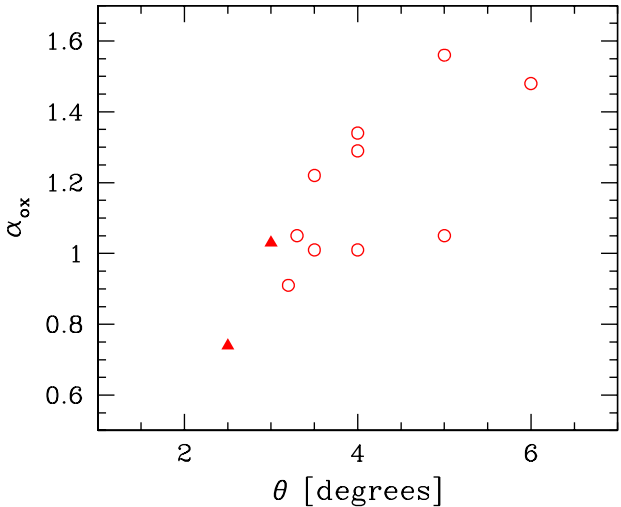,height=10.3cm,width=10.3cm}}
\vspace*{-0.8 truecm}
\caption{The broad band spectral UV to X--ray spectral index
 $\alpha_{OX}$ as a function of viewing angle.  Filled points are the
 high--$z$ blazars (triangles).}
\label{a_ox_theta}
\end{figure}

We find from the modelling that $\alpha_{ox}$ correlates
with the viewing angle $\theta$ (Fig.~\ref{a_ox_theta}) as well as
(inversely) with the Doppler factor $\delta$ (not shown). 
We recall that $\Gamma$ and $\delta$ enter the models in different 
ways so that both are determined in the fits: $\theta$ is then derived 
from the first two.    
The two correlations are not independent as the value of $\Gamma$ is 
almost constant ranging between 10 and 15.

The $\alpha_{ox}$ -$\theta$ correlation agrees with the initial expectation 
that relatively weak jet contributions could derive from intermediate
viewing angles. However $\alpha_{ox}$ also
correlates {\it inversely} with the estimated jet intrinsic power,
so that the two effects,
a lower degree of beaming or a lower intrinsic jet power, cannot be 
disentangled and probably coexist.
On the other hand, $\alpha_{ox}$ does not correlate with $L_{disk}$,
indicating that it is the {\it relative} jet
"strength", either due to the viewing angle or to the intrinsic power, 
that may drive these correlations.

\begin{figure}
\centering
\hspace*{-1.1 cm}
\includegraphics[scale=0.5]{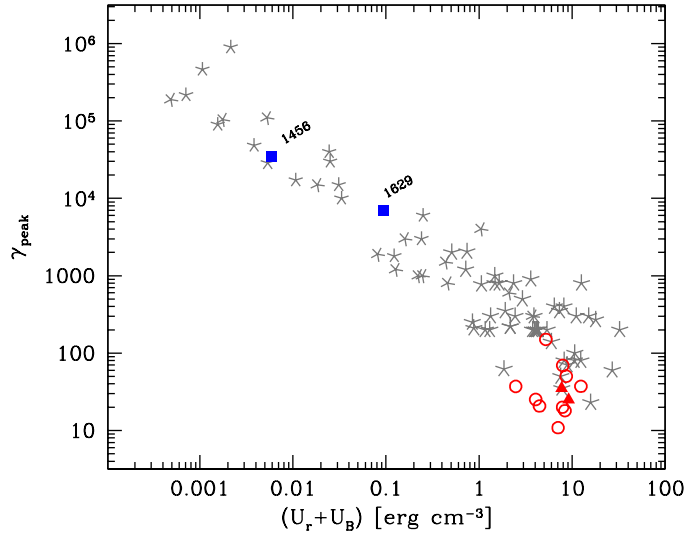}
\caption{The energy of electrons emitting at the peaks of the SED,
$\gamma_{\rm peak}$ as a function of the energy density (radiative
plus magnetic) as seen in the comoving frame. Oue EMSS sources are
shown as empty circles. The two squares correspond to the two ``blue''
FSRQs in our sample, while the two filled triangles correspond to the
two high redshift blazars SDSS~J$081009.94+384757.0$ and
MG3~J$225155+2217$. For comparison, we show also all the blazars
studied in Celotti \& Ghisellini (2008) (grey asterisks).}
\label{gpeak}
\end{figure}

The most important parameter determining the {\it shape} of
the SED is $\gamma_{\rm peak}$, the energy of electrons radiating the
peak synchrotron luminosity. The relation between $\gamma_{\rm peak}$
and the radiation energy density as seen in the emission region frame, $u_{rad}$,
 is shown in Fig.~\ref{gpeak}, for the
objects discussed here.  Values derived for the blazars modelled in
CG08 are also shown for comparison.  The new objects fall well within
the parameter sequence defined by the previous sample. Therefore,
despite the different selection criteria, X-ray selection as opposed
to radio selection for FSRQ  and selections aimed at finding
sources breaking the sequence trends, the spectral sequence holds in
terms of physical parameters.  


The ``red'' objects tend to cluster at one end, with rather low 
$\gamma_{\rm peak}$.  The clustering can result from the assumption 
(adapted from Kaspi et al. 2007) of an almost constant intrinsic 
radiation energy density in the BLR: since the magnetic field value 
is limited by the high Compton to synchrotron ratio, 
$U_{\rm rad} \simeq U_{BLR} \Gamma^2$ (which varies by a factor 2 
for $\Gamma$ between 10 and 15) cannot be exceeded substantially.

Blue SEDs are obtained for small $U_{\rm rad}$, such that $\gamma_{\rm
peak} = \gamma_{\rm c}$ is high, while the opposite is true for red
SEDs. This is a result of the feed-back introduced by including
radiative cooling in modelling the energy distribution of the
relativistic electrons.  However this is not the only important
difference between the electron distributions producing blue and red
SEDs.  Another major difference is at the lower electron energies: in fact,
below $\gamma_{\rm peak}$, the assumed electron distribution is
extremely hard ($\propto \gamma^{-1}$) so that the total number of
electrons is small and their average energy much higher than in the
case of red SEDs (see Table 6).

\subsection {Jet powers}

\begin{figure}
\centering
{\hspace*{-1.1 truecm}
\psfig{file=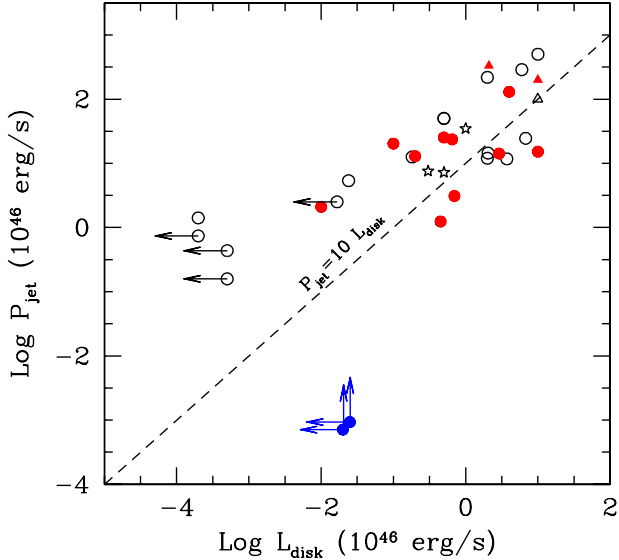,height=10.5cm,width=10.5cm}}
\vspace*{-0.9 truecm}
\caption{The total jet power as a function of the disk luminosity, for
the blazars studied in this paper. Filled red triangles are the
high--$z$ blazars and filled blue circles are the ``blue'' blazars. For
comparison, black open symbols report the data obtained for other blazars
(Maraschi \& Tavecchio 2003, Sambruna et al 2006b).
}
\label{table}
\end{figure}

The total jet power, $P_{\rm jet}$, computed from the models (see
Sect.\ref{modelling}) is shown vs. the luminosity of the accretion
disk (estimated from the data) in Fig.~\ref{table}, together with
values previously derived for other blazars (Maraschi \& Tavecchio
2003, Sambruna et al. 2006b).  The two quantities tend to correlate
for the red SEDs. Moreover the value of $P_{\rm jet}$ is of the order
of 10 times the disk luminosity, corresponding to an approximate equality 
of the jet power and the accretion power (for an assumed radiative
efficiency of the accretion process of ~ 10\% appropriate for an
optically thick flow).  This represents on one hand an interesting consistency
check on the models. In fact the bulk Lorentz factor, size, magnetic
field, particle densities and spectra in the dissipation region are
chosen to satisfy the SED constraints, including clearly the observed
fluxes/luminosities, but the jet power is a global quantity that is
computed \emph{a posteriori}. Moreover the new data and models
increase significantly the number of sources in which a comparison 
of the jet power with the accretion power is possible and confirms
the substantial balance between the two.

The value of the power for jets with red SEDs depends quite
strongly on the minimum energy of the injected electrons, $\gamma_{\rm
min}$, which is often poorly constrained.  For red SEDs the X-ray
emission derives from the EC process of relatively low energy
electrons thus a lower limit can be obtained. Moreover in several
cases, $\gamma_{\rm min}$ could be inferred from detailed fits of
X-ray spectra showing a break in the X-ray range. Such breaks have
sometimes been interpreted as absorption by cold gas (e.g. Cappi et
al. 1997, Fiore et al. 1998, Fabian et al. 2001a,b, Bassett et
al. 2004) but could represent instead the low energy "end" of the
electron distribution (e.g. Tavecchio et al. 2007). With the latter
interpretation $\gamma_{\rm min}$ is found to be in the range 1-10,
thus supporting the high power values (Tavecchio et al. 2000,
Maraschi \& Tavecchio 2003, CG08). These powers rest on the assumption
of 1 cold proton per electron, but are independently required by the
fact that the power directly radiated by the jet is larger than  the
power in relativistic leptons and magnetic field (CG08).

For jets with blue SEDs the jet power derived from the present models
is substantially lower, due to the smaller number and higher mean
energy of particles in the assumed relativistic electron population.
The uncertainty on the amount of low energy electrons in the jet is
however quite large because radiation from this branch of the electron
distribution is not observable: its synchrotron emission is covered by
the larger fluxes due to the outer regions of the jet and its SSC
emission falls in the hard X-ray - soft gamma-ray region of the
spectrum which is difficult to observe.  Note that from Table 6 the
total kinetic power from the "blue" jets is hardly sufficient to
account for the emitted radiation, thus the assumed spectra lead to a
lower limit in the estimate of the jet powers.

As a result, in the $P_{\rm jet}$ -- $L_{disk}$ plane, the "blue"
objects modelled here fall well below the $P_{\rm jet}$ = 10
$L_{disk}$ line. In the same diagram the grey open circles with
upper limits to the disk luminosities (horizontal arrows) represent BL
Lacs from Maraschi \& Tavecchio (2003) for which an electron
distribution with slope 2 extending down to a Lorentz factor of 1 was
assumed below the peak energy.

Despite the large uncertainties about the powers of the "blue"
objects, it is interesting that the "red" blazars discussed here which
result from independent selection criteria in the X-ray/hard X-ray band
confirm the results previously obtained for radioselected FSRQ.

Very different results were obtained by Nieppola et al.  (2008)
for a large sample of Blazars using beaming corrections derived from
brightness temperatures at high radio frequencies using variability to
estimate the sizes of the emission regions. This technique gives
information on the Doppler factors for the radioemitting regions in
AGN. They find an anticorrelation between Doppler factor and peak
frequency of the synchrotron component such that the objects with
higher peak frequency are less beamed.  As a result the "intrinsic"
powers of objects with large peak frequency (i.e. "blue" objects) are
higher than those of "red" objects. 

While models of TeV emitting blazars require Doppler factors as
large as 10-50 (e.g. Krawczynski et al. 2002, Konopelko et al. 2003,
Finke et al. 2008, Aharonian et al. 2008, Tagliaferri et al. 2008),
VLBI observations show very small, subluminal motions (e.g. Piner,
Pant \& Edwards 2008). These two evidences can be reconciled assuming
that the jet strongly decelerates from the blazar region to the VLBI
scale, from which most of the radio emission originates
(Georganopoulos et al. 2003, Ghisellini et al. 2005). This would also
fit in with the scenario in which FR I jets decelerate on moderate
scale lengths while FR II jets may remain relativistic up to very
large scales (e.g. Tavecchio 2007). Therefore the Doppler factors
obtained by Nieppola et al. (2008) for the jet radio emitting regions
are probably much smaller than the Doppler factors in the
optical-to-$\gamma$-ray emitting region especially in the case of high
peaked BL Lac objects.

\begin{figure*}
\vspace*{-1.9 truecm}
{\hspace*{-1.1 truecm}
\psfig{file=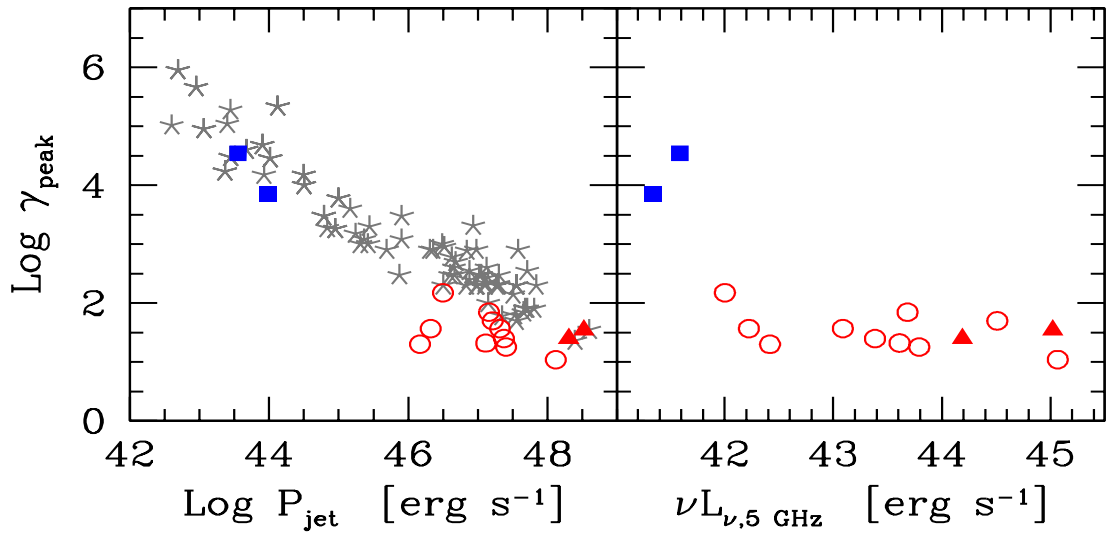,height=15cm,width=17cm}}
{\vspace*{-3.9 truecm}
\caption{The energy of electrons emitting at the peaks of the SED,
$\gamma_{\rm peak}$ as a function of total jet power ({\it left}, for
comparison in gray data from CG08) and the radio luminosity at 5 GHz
({\it right}). Symbols as in Fig.9. A ``sequence'' (i.e. correlation) is still apparent
between $\gamma_{\rm peak}$ and $P_{\rm jet}$, while $\gamma_{\rm
peak}$ does not show a clear correlation with observed radio
luminosity.}}
\label{gpeakl}
\end{figure*}

\section{Conclusions}

We have analysed and modelled new data for the SEDs of blazars
including the first X-ray selected sample of Radio-loud Quasars and a few more
blazars claimed to possibly challenge the ``blazar spectral
sequence''.  The first, model independent, conclusion is that the
conventional separation of different ``types'' of blazars according to
their two point spectral indices ($\alpha_{ro}, \alpha_{ox},
\alpha_{rx}$) can be deceiving, due to the complexity and variety of
the SEDs: for instance the traditional criterion $\alpha_{rx}>(<)
0.78$ to distinguish between blue and red SEDs fails in the case of
SDSS~J$081009.94+384757.0$ and MG3~J$225155+2217$ which have 
$\alpha_{rx} \simeq 0.6$, yet this is not due to synchrotron emission in the
X-ray band but to an extremely dominant 
Compton component. Results from the Fermi Gamma-ray Telescope will be critical 
in confirming our interpretation

The ``blazar spectral sequence'' concept still holds in terms of a
parameter sequence. In Fig.11 (left panel) the parameter $\gamma_{\rm
peak}$ computed from SED models is plotted vs. total jet power also
computed from the models. A ``sequence'' (i.e. correlation) is still
apparent between these two quantities, We stress that $\gamma_{\rm
peak}$ is related to the spectral shape while $P_{jet}$ is a global
quantity related to the emitted luminosity.

However (Fig.11, right panel) $\gamma_{\rm peak}$ does not show a
clear correlation with the observed radio luminosity, which was the
basic quantity used in building the spectral sequence. This is likely
due to the fact that at lower flux thresholds intrinsically luminous
but less beamed objects enter the samples causing a mix of lower
luminosity due to lower intrinsic power or lower beaming. This
ambiguity could probably be solved with high resolution radio
observations.

The spectral sequence requires that different ``types''
of blazars, from FSRQ to low--energy peaked BL Lac Objects (LBLs) to
HBLs, have SED peaks at increasing frequencies. This is accounted for
by increasing values of $\gamma_{\rm peak}$, derived in our models
with the assumption that radiative cooling affects the electron
distribution becoming less important as the radiation energy density
around the dissipation region decreases.  However the particle
distributions needed to account for red and blue SEDs respectively
differ not only at the high energy end but also at the low energy
end. Blue SEDs require an electron population with higher "average"
energy, perhaps pointing to different injection/acceleration
mechanisms.

The present results show that even X-ray selection does not lead to
find "blue" quasars as was the case for X-ray selected BL Lac samples
(see also Landt et al. 2008).  This strengthens the conclusion that
when a bright accretion disk is present the SED is always "red".
However, if the initial samples suggested luminosity/power as a
fundamental parameter in determining the SEDs' systematics, the
present study suggests that the fundamental condition that separates
``red'' SEDs from ``blue'' ones is the intensity of the radiation
field in the emission region, which is dominated by reprocessed
photons from an accretion disk, if present. The relation with
luminosity/power could be indirect, due to the fact that a bright
accretion disk forms only for high values of the accretion rate in
Eddington units (e.g. $\dot{m} > 10^{-2} \dot{m}_{\rm crit}$), while
the absence of it (RIAF, Radiatively Inefficient Accretion Flows)
occurs for $\dot{m} < 10^{-2} \dot{m}_{\rm crit}$, thus at lower power
{\it for a fixed mass of the accreting Black Hole}.  Thus, the
original picture of blazar SEDs, as a one parameter family governed by
luminosity, should be revised including the mass of the accreting
Black Hole as an additional important parameter.  A new scheme along
these lines has been recently suggested by Ghisellini \& Tavecchio
(2008).

\section*{Acknowledgements} 
We thank an anonymous referee for useful comments.  This research has
made use of data obtained from the High Energy Astrophysics Science
Archive Research Center (HEASARC), provided by NASA's Goddard Space
Flight Center.

This research has made use of the NASA/IPAC Extragalactic Database
(NED) which is operated by the Jet Propulsion Laboratory, California
Institute of Technology, under contract with the National Aeronautics
and Space Administration.

Funding for the SDSS and SDSS-II has been provided by the Alfred
P. Sloan Foundation, the Participating Institutions, the National
Science Foundation, the U.S.  Department of Energy, the National
Aeronautics and Space Administration, the Japanese Monbukagakusho, the
Max Planck Society, and the Higher Education Funding Council for
England. The SDSS Web Site is http://www.sdss.org/.  The SDSS is
managed by the Astrophysical Research Consortium for the Participating
Institutions. The Participating Institutions are the American Museum
of Natural History, Astrophysical Institute Potsdam, University of
Basel, University of Cambridge, Case Western Reserve University,
University of Chicago, Drexel University, Fermilab, the Institute for
Advanced Study, the Japan Participation Group, Johns Hopkins
University, the Joint Institute for Nuclear Astrophysics, the Kavli
Institute for Particle Astrophysics and Cosmology, the Korean
Scientist Group, the Chinese Academy of Sciences (LAMOST), Los Alamos
National Laboratory, the Max-Planck-Institute for Astronomy (MPIA),
the Max-Planck-Institute for Astrophysics (MPA), New Mexico State
University, Ohio State University, University of Pittsburgh,
University of Portsmouth, Princeton University, the United States
Naval Observatory, and the University of Washington.

We acknowledge funding from ASI/INAF with contract I/088/06/0.

\clearpage

\begin{table*}
	\centering
\caption{\emph{Swift} observation log.}	
		\begin{tabular}{lcc}
			Source & ObsID & Date\\
			\hline
			MS $0152.4+0424$  & $00036502001$ & $28-06-2007$ \\
			{}                & $00036502002$ & $11-01-2008$ \\
			MS $0232.5-0414$  & $00036503001$ & $21-02-2008$ \\
			{}                & $00036503002$ & $22-02-2008$ \\
			MS $0402.0-3613$  & $00035523001$ & $24-03-2006$ \\			
			{}                & $00035523002$ & $29-03-2006$ \\
			{}                & $00036504001$ & $27-07-2007$ \\
			{}                & $00036504002$ & $08-08-2007$ \\			
			{}                & $00036504003$ & $30-12-2007$ \\
			{}                & $00036504004$ & $01-01-2008$ \\
			MS $0521.7+7918$  & $00036505001$ & $01-12-2007$ \\
			{}                & $00036505002$ & $02-12-2007$ \\
			{}                & $00036505003$ & $04-12-2007$ \\			
			MS $0828.7+6601$  & $00036506001$ & $23-01-2008$ \\
			MS $1050.9+5418$  & $00036507001$ & $24-01-2008$ \\			
			{}                & $00036507002$ & $27-01-2008$ \\
			MS $1234.9+6651$	& $00036508001$ & $18-05-2007$ \\		
			{}								& $00036508002$ & $12-07-2007$ \\			
			{}								& $00036508003$ & $09-10-2007$ \\			
			{}								& $00036508004$ & $10-10-2007$ \\			
			{}								& $00036508005$ & $11-10-2007$ \\												
			MS $1340.7+2859$  & $00036509001$ & $30-05-2007$ \\			
			{}  							& $00036509002$ & $01-07-2007$ \\			
			MS $1623.4+2712$	& $00036510001$ & $14-11-2007$ \\		
			{}								& $00036510002$ & $17-11-2007$ \\					
			MS $2141.2+1730$  & $00036358001$ & $19-04-2007$ \\
			{}  							& $00036358002$ & $23-04-2007$ \\			
			{}                & $00036511001$ & $09-08-2007$ \\						
			{}                & $00036511004$ & $16-10-2007$ \\						
			{}                & $00036511005$ & $24-12-2007$ \\						
			{}                & $00036511006$ & $11-01-2008$ \\						
			{}                & $00036511007$ & $12-01-2008$ \\															
			{}                & $00036511008$ & $13-01-2008$ \\						
			\hline
			SDSS J$081009.94+384757.0$ & $00030370001$ & $03-03-2006$\\
			{} 												 & $00036229001$ & $21-12-2007$\\			
			MG3 J$225155+2217$         & $00037099001$ & $16-05-2007$\\
			{}         								 & $00037099002$ & $21-05-2007$\\						
			{}         								 & $00037099003$ & $22-05-2007$\\									
			{}         								 & $00036660001$ & $26-05-2007$\\									
			\hline
			RX J$1456.0+5048$          & $00030925001$ & $27-04-2007$ \\
			RGB J$1629+401$            & $00035022001$ & $20-04-2005$\\
			{}                         & $00035022002$ & $23-05-2005$\\			
			{}                         & $00035400001$ & $20-01-2006$\\						
			{}                         & $00035022003$ & $22-04-2007$\\									
			{}                         & $00035022004$ & $28-04-2007$\\									
			{}                         & $00035022005$ & $04-05-2007$\\									
			{}                         & $00035022006$ & $12-05-2007$\\									
			{}                         & $00035022007$ & $18-05-2007$\\									
			{}                         & $00035022008$ & $27-05-2007$\\
			{}                         & $00036549001$ & $26-06-2007$\\
			{}                         & $00036549002$ & $08-01-2008$\\
			{}                         & $00036549003$ & $11-01-2008$\\						
 			\hline
		\end{tabular}
\label{table:obslog1}
\end{table*}

\begin{table*}
	\centering
\caption{\emph{Swift}/XRT measurements. $N_{\rm H}$ (from Kalberla et al. 2005) and $N_{\rm H}^{z}$ are in units of $10^{20}$~cm$^{-2}$; exposure is in kiloseconds; $\Gamma_1$ indicates the photon index in the case of a single power-law model or the soft photon index in the case of a broken power-law model; for the latter, the columns $\Gamma_2$ and $E_{\rm break}$ indicates the hard photon index and the break energy [keV]; the flux is in units of $10^{-13}$~erg~cm$^{-2}$~s$^{-1}$.}	
		\begin{tabular}{lccccccccc}
			\hline
			Source & $z$ & $N_{\rm H}$ & Exposure & $N_{\rm H}^{z}$ & $\Gamma_1$ & $\Gamma_2$ & $E_{\rm break}$ & $\tilde{\chi}^2$/dof & $F_{2-10 \ \rm keV}$\\
 			\hline
			MS $0152.4+0424$  & $1.132$ & $4.00$  & $10.0$ & {} & $1.5\pm 0.2$ & {} & {} & $0.58/8$ & $9.64$\\
			MS $0232.5-0414$  & $1.439$  & $2.26$  & $15.4$ & {} & $1.72\pm 0.08$ & {} & {} & $0.93/25$ & $16.3$\\
			MS $0402.0-3613$  & $1.417$ & $0.603$ & $27.7$ & $20\pm 10$ &$1.76\pm 0.09$ & {} & {} & $1.20/29$ & $18.3$\\
			MS $0521.7+7918$  & $0.503$ & $6.93$  & $21.5$ & {} &$1.9\pm 0.2$ & {} & {} & $0.53/6$ & $3.70$\\
			MS $0828.7+6601$  & $0.329$ & $4.31$  & $9.3$  & {} &$1.3\pm 0.2$ & {} & {} & $0.12/5$ & $9.20$\\
			MS $1050.9+5418$  & $0.995$ & $0.891$ & $11.2$ & {} &$1.8\pm 0.4$ & {} & {} & $0.86/4$ & $3.56$\\
			MS $1234.9+6651$	& $0.852$ & $1.75$  & $32.8$ & {} &$1.8\pm 0.1$ & {} & {} & $1.37/19$ & $5.49$\\
			MS $1340.7+2859$  & $0.905$ & $1.24$  & $12.1$ & {} &$1.6\pm 0.1$ & {} & {} & $0.42/12$ & $11.8$\\
			MS $1623.4+2712$	& $0.526$ & $3.25$  & $15.6$ & {} &$1.8\pm 0.2$ & {} & {} & $0.63/10$ & $7.60$\\
			MS $2141.2+1730$  & $0.211$ & $7.35$  & $43.0$ & {} &$1.69\pm 0.06$ & {} & {} & $1.67/35$ & $15.9$\\
 			\hline
			SDSS J$081009.94+384757.0$ & $3.946$ & $4.88$  & $10.6$ & {} & $1.4\pm 0.3$ & {} & {} & $0.28/3$ & $5.39$\\
			MG3 J$225155+2217$ & $3.668$ & $4.90$  & $23.1$ & $210_{-110}^{+140}$ & $1.41_{-0.09}^{+0.10}$ & {} & {} & $1.05/42$ & $31.0$\\
			\hline
			RX J$1456.0+5048$ & $0.478$ & $1.60$  & $4.8$ & {} & $2.2\pm 0.1$ & {} & {} & $0.77/25$ & $17.7$\\
			RGB J$1629+401$ & $0.272$ & $0.977$  & $40.7$ & {} & $3.4_{-0.3}^{+0.7}$ & $2.21\pm 0.06$ & $0.47\pm_{-0.05}^{+0.08}$ & $1.15/78$ & $9.36$\\
			\hline 			
		\end{tabular}
\label{table:xrt}
\end{table*}

\begin{table*}
	\centering
	\caption{\emph{Swift}/BAT and \emph{INTEGRAL}/ISGRI measurements. Exposures are in kiloseconds; fluxes are in units of $10^{-10}$~erg~cm$^{-2}$~s$^{-1}$; upper limits are at $3\sigma$ level, after having taken into account systematic errors. The Crab spectrum taken as reference for conversion in physical units can be expressed with a power-law model with $\Gamma = 2.1$ and normalization at 1 keV equal to $9.7$~ph~cm$^{-2}$~s$^{-1}$~keV$^{-1}$ (Toor \& Seward 1974).}
		\begin{tabular}{lccc}
			\hline
			Source & Exposure & $F_{20-40 \ \rm keV}$ & $F_{40-100 \ \rm keV}$\\
 			\hline
			MS $0152.4+0424$  & $10.0$ & $<2.1$ & $<3.8$\\
			MS $0232.5-0414$  & $15.9$ & $<2.8$ & $<3.5$\\
			MS $0402.0-3613$  & $28.5$ & $<1.7$ & $<2.2$\\			
			MS $0521.7+7918$  & $22.0$ & $<2.4$ & $<3.1$\\
			MS $0828.7+6601$  & $9.6$  & $<3.2$ & $<4.0$\\
			MS $1050.9+5418$  & $11.6$ & $<2.6$ & $<3.5$\\			
			MS $1234.9+6651$	& $32.7$ & $<1.5$ & $<2.1$\\		
			MS $1340.7+2859$  & $12.4$ & $<2.7$ & $<3.9$\\			
			MS $1623.4+2712$	& $16.0$ & $<2.5$ & $<3.6$\\		
			MS $2141.2+1730$  & $44.8$ & $<1.8$ & $<2.2$\\
 			\hline
			SDSS J$081009.94+384757.0$ & $11.0$ & $<2.4$ & $<3.8$ \\
			MG3 J$225155+2217$			   & $26.8$ & $<2.3$ & $<2.9$\\
			MG3 J$225155+2217^*$       & $388$  & $0.12\pm 0.03$ & $0.32\pm 0.07$\\
			\hline
			RX J$1456.0+5048$          & $5.0$  & $<3.3$ & $<4.9$ \\
			RGB J$1629+401$						 & $42.8$ & $<1.3$ & $<2.0$ \\			
			\hline 			
		\end{tabular}
\begin{list}{}{}
\item[$^{*}$] \emph{INTEGRAL} observation.
\end{list}
\label{table:hardx}
\end{table*}

\begin{table*}
	\centering
\caption{\emph{Swift}/UVOT magnitudes (lower limits are at $3\sigma$ level) and the corresponding 
broad band spectral indices $\alpha_{\rm ro}$, $\alpha_{\rm ox}$ and $\alpha_{\rm rx}$. These have been calculated by using the K--corrected fluxes at ($3501$~\AA), $1$~keV and $5$~GHz. See Sect.~3.3 for more details.}	
\begin{tabular}{lccccccccc}
\hline
Source & V & B & U & UVW1 & UVM2 & UVW2 &$\alpha_{\rm ro}$  &$\alpha_{\rm ox}$ &$\alpha_{\rm rx}$   \\
\hline
MS $0152.4+0424$  &$18.6\pm 0.2$ &$19.2\pm 0.2$ &$18.5\pm 0.1$ &{} & {} & {} &0.61 &$1.05$  &0.75 \\
MS $0232.5-0414$  &$16.6\pm 0.1$ &$16.7\pm 0.1$ &$15.5\pm 0.1$ &{} & {} & {} &0.43 &$1.29$  &0.70 \\
MS $0402.0-3613$  &$17.5\pm 0.1$ &$17.7\pm 0.1$ &$16.6\pm 0.1$ &{} & {} & {} &0.60 &$1.01$  &0.74 \\			
MS $0521.7+7918$  &$17.1\pm 0.1$ &$17.4\pm 0.1$ &$16.6\pm 0.1$ &{} & {} & {} &0.38 &$1.48$  &0.73 \\
MS $0828.7+6601$  &$>19.7$       &$>20.5$       &$>20.3$       &{} & {} & {} &$<0.77$ &$<0.91$ &0.81 \\
MS $1050.9+5418$  &$19.3\pm 0.3$ &$19.8\pm 0.3$ &$18.7\pm 0.1$ &{} & {} & {} &0.57 &$1.01$  &0.71  \\			
MS $1234.9+6651$  &$17.7\pm 0.1$ &$18.2\pm 0.1$ &$17.2\pm 0.1$ &{} & {} & {} &0.48 &$1.22$  &0.71 \\		
MS $1340.7+2859$  &$16.7\pm 0.1$ &$17.1\pm 0.1$ &$16.0\pm 0.1$ &{} & {} & {} &0.39 &$1.34$  &0.69\\			
MS $1623.4+2712$  &$18.8\pm 0.2$ &$18.9\pm 0.1$ &$18.2\pm 0.2$ &{} & {} & {} &0.60 &$1.05$  &0.75  \\		
MS $2141.2+1730$  &$16.2\pm 0.1$ &$16.4\pm 0.1$ &$15.2\pm 0.1$ &{} & $15.1\pm 0.1$ &$15.0\pm 0.1$ &0.49 &$1.56$ &0.83 \\
\hline
SDSS J$081009.94+384757.0$ 
                  &$>18.5$       &$>19.4$       &$>19.2$       &$>18.7$       &$>18.9$ & $>19.5$            &$<0.34$ &$<1.03$ &0.56 \\
MG3 J$225155+2217$&{}            &{}            &$19.3\pm 0.1$ &$18.9\pm 0.2$ &$18.4\pm 0.1$ &$18.5\pm 0.1$ &0.52    &$0.74$  &0.59 \\
\hline
RX J$1456.0+5048$ &$18.6\pm 0.3$ &$19.2\pm 0.2$ &$18.4\pm 0.2$ &$17.9\pm 0.2$ &$17.9\pm 0.2$ &$17.9\pm 0.1$ &0.42 &$0.69$ &0.55 \\
RGB J$1629+401$   &$17.9\pm 0.2$ &$18.2\pm 0.2$ &$17.2\pm 0.1$ &$16.8\pm 0.1$ &$16.5\pm 0.1$ &$16.5\pm 0.1$ &0.35 &$1.26$ &0.66 \\
\hline 			
\end{tabular}
\label{table:uvot}
\end{table*}

\clearpage

\begin{table*}
\caption{The input parameters of the model for our blazars.
(1) Source name; 
(2) redshift;
(3) radius $R$ of emitting region [$10^{15}$~cm]; 
(4) intrinsic injected power [$10^{45}$~erg/s];
(5) bulk Lorentz factor;
(6) viewing angle;
(7) magnetic field intensity [Gauss]; 
(8) minimum random Lorentz factor of the injected particles; 
(9) maximum random Lorentz factor of the injected particles; 
(10) $\gamma_{\rm peak}$; 
(11) spectral slope of particles above the cooling break; 
(12) disk luminosity [$10^{45}$~erg/s];
(13) radius of the BLR [$10^{15}$~cm].
(14) random Lorentz factor of the electrons cooling in $\Delta R^\prime/c$.
}
\begin{tabular}{@{}llccccccccccccl}
\hline 
Source &$z$ &$R$ &$L^\prime_{\rm inj}$ &$\Gamma$ &$\theta$ &$B$ &$\gamma_{\rm inj}$ 
&$\gamma_{\rm max}$ &$\gamma_{\rm peak}$ &$n$ &$L_{\rm d}$ &$R_{\rm BLR}$ &$\gamma_{\rm c}$  &Note \\ 
(1) &(2) &(3) &(4) &(5) &(6) &(7) &(8) &(9) &(10) &(11) &(12) &(13) &(14)  &(15)\\ 
\hline
MS 0152.4+0424  &1.132 &10 &1.5e-2 &14 &3.3 &2.4  &1    &3e3   &18    &3.2  &5     &220   &11   & \\
MS 0232.5-0414  &1.439 &18 &3.0e-2 &13 &4.0 &7    &1    &6e3   &50    &3.3  &100   &1.e3  &6    & \\
MS 0402.0-3613  &1.417 &12 &2.3e-2 &11 &4.0 &8    &4    &1e4   &11    &3.4  &40    &700   &11   &\\ 
MS 0521.7+7918  &0.503 &10 &1.0e-2 &13 &6.0 &2    &150  &8e3   &150   &3.5  &7     &300   &18   &\\
MS 0828.7+6601  &0.329 &2  &6.0e-5 &14 &3.2 &3    &1    &1e4   &37    &3.45 &0.1   &25    &37   &  \\ 
MS 1050.9+5418  &0.995 &10 &1.5e-3 &12 &3.5 &5    &1    &1e4   &21    &3.2  &2     &180   &21   &$R_{\rm diss}<R_{\rm BLR}$ \\
                  &      &18 &2.5e-3 &12 &3.5 &0.12 &400  &2e5   &7e3   &3.5  &2     &140   &7e3  &$R_{\rm diss}>R_{\rm BLR}$ \\
MS 1234.9+6651  &0.852 &9  &7.0e-3 &12 &3.5 &3    &15   &5e3   &25    &3.6  &6.5   &320   &25   & \\
MS 1340.7+2859  &0.905 &14 &2.5e-2 &13 &4.0 &3    &70   &3e3   &70    &3.6  &29    &500   &8    &$L_\gamma$ below GLAST\\
                  &      &14 &2.0e-2 &14 &4.5 &5    &40   &3e3   &40    &3.7  &25    &550   &9    &$L_\gamma$ above GLAST\\  
MS 1623.4+2712  &0.526 &10 &1.6e-3 &11 &5.0 &3    &1    &1e4   &37    &3.3  &1     &150   &37   &$R_{\rm diss}<R_{\rm BLR}$ \\ 
                  &      &15 &3.6e-4 &11 &3.5 &0.1  &300  &1e4   &2.1e4 &3.6  &1     &100   &2e4&$R_{\rm diss}>R_{\rm BLR}$ \\ 
MS 2141.2+1739  &0.211 &10 &9.0e-4 &12 &5.0 &6.5  &20   &1.5e4 &20    &2.4  &4.5   &200   &12   &\\
\hline
SDSSJ081009.94+384757.0 
                  &3.946 &16 &0.1    &14 &3.0 &5     &25   &1.5e3 &25    &3.8 &100   &1000  &6    &\\
MG3 J225155+2217&3.668 &20 &0.2    &15 &2.5 &0.7   &35   &1e3   &35    &3.8 &21    &500   &6    & \\
\hline 
RX J1456.0+5048 &0.478 &15 &2.0e-4 &15 &3.0 &0.3  &3.5e4 &5e5   &3.5e4 &3.4 &0.2   &45   &1e4  &$R_{\rm diss}>R_{\rm BLR}$ \\
RGB J1629+401   &0.272 &10 &1.1e-4 &10 &4.0 &1.5  &7e3   &1e5   &7e3   &3.5 &0.25  &50   &930  &$R_{\rm diss}>R_{\rm BLR}$ \\
\hline
\hline
\end{tabular}
\label{table:para}
\end{table*}

\begin{table*}
\caption{Kinetic powers and Poynting fluxes 
(all in units of $10^{45}$~erg/s) ---
(1) Source name; 
(2) Total (synchrotron + IC) radiative power $L_{\rm r}$;
(3) Synchrotron radiative power $L_{\rm s}$;
(4) Poynting flux $L_{\rm B}$;
(5) Kinetic power in emitting electrons $L_{\rm e}$;
(6) Kinetic power in protons $L_{\rm p}$, assuming one proton per electron;
(7) Average random electron Lorentz factor $\langle\gamma\rangle$.
}
\begin{tabular}{@{}llllllll}
\hline 
Source &$L_{\rm r}$   &$L_{\rm s}$  &$L_{\rm B}$  &$L_{\rm e}$  &$L_{\rm p}$  &$\langle\gamma\rangle$ &Note \\
(1) &(2) &(3) &(4) &(5) &(6) &(7) &(8)\\ 
\hline
MS $0152.4+0424$   &3.15    &0.04    &0.42   &1.11     &253      &8.1   &  \\
MS $0232.5-0414$   &4.92    &0.73    &10.03  &0.68     &152      &8.2   & \\
MS $0402.0-3613$   &2.29    &0.33    &4.16   &2.90     &1301     &4.1   & \\  
MS $0521.7+7918$   &1.73    &0.04    &0.25   &0.29     &31       &17.6  & \\  
MS $0828.7+6601$   &1.20e-2 &1.53e-4 &2.64e-2 &3.24e-2 &21       &2.9   &   \\  
MS $1050.9+5418$   &0.20    &2.67e-2 &1.25   &0.24     &129      &3.4   &$R_{\rm diss}<R_{\rm BLR}$  \\
                   &0.39    &2.90e-2 &2.51e-3 &1.27    &14.4     &162   &$R_{\rm diss}>R_{\rm BLR}$  \\  
MS $1234.9+6651$   &1.02    &2.71e-2 &0.39   &1.10     &236      &8.6   &  \\  
MS $1340.7+2859$   &4.32    &9.73e-2 &1.11   &0.78     &142      &10    &$L_\gamma$ below GLAST  \\  
                   &3.80    &0.195   &3.59   &1.22     &254      &8.8   &$L_\gamma$ above GLAST  \\  
MS $1623.4+2712$   &0.19    &1.50e-2 &0.41   &3.66     &203      &3.3   &$R_{\rm diss}<R_{\rm BLR}$  \\   
                   &5.60e-2 &6.0e-3  &1.0e-3 &0.96     &15       &117   &$R_{\rm diss}>R_{\rm BLR}$  \\ 
MS $2141.2+1739$   &0.12    &1.38e-2 &2.27   &5.52e-2  &12.4     &8.2   & \\  
\hline
SDSS J$081009.94+384757.0$  
                   &19.79   &0.44    &4.69    &7.10     &1996     &6.5  & \\
MG3 J$225155+2217$ &50.05   &2.12e-2 &0.16    &12.77    &3320     &7    & \\ 
\hline
RX J$1456.0+5048$  &3.26e-2 &2.13e-2 &1.70e-2 &1.17e-2  &7.1e-3   &3049 &$R_{\rm diss}>R_{\rm BLR}$ \\
RGB J$1629+401$    &1.13e-2 &1.03e-2 &8.39e-2 &2.23e-3  &9.3e-3   &442  &$R_{\rm diss}>R_{\rm BLR}$\\
\hline
\hline
\end{tabular}
\label{table:power}
\end{table*}

\end{document}